**Optimization of Magnetic Milli-Spinner for Robotic Endovascular Intervention**

*Lu Lu, Luca Higgins, Jack Bernardo, Ruike Renee Zhao**

L. Lu, L. Higgins, J. Bernardo, R. R. Zhao
Department of Mechanical Engineering, Stanford University, Stanford, CA 94305, United States
Email: rrzhao@stanford.edu (R.R. Zhao)

## Abstract

Vascular diseases such as atherosclerosis, thrombosis, and aneurysms can lead to life-threatening medical events. Conventional catheter- or guidewire-based interventional devices often struggle to navigate through highly tortuous vasculature. The recently developed multifunctional magnetic milli-spinner offers a promising wireless solution by integrating a central through-hole and side slits into a cylindrical body with helical fins, enabling rapid and stable navigation for clot debulking, targeted drug delivery, and aneurysm treatment. Here, we combine computational fluid dynamics simulations with experimental validation to optimize the milli-spinner's structural design for high-velocity propulsion and high-efficiency clot debulking in tubular flow environments. By systematically investigating the effects of through-hole radius, fin number, fin helical angle, and slit dimension on propulsion performance, the optimized milli-spinner achieves swimming velocities of 55 cm/s (~175 body lengths/s ) in saline water and 44 cm/s (~140 body lengths/s) in a fluid with viscosity (3.5 mPa·s) comparable to that of arterial blood at high shear rates, far exceeding existing untethered magnetic robots in tubular environments (< 80 body lengths/s). This exceptional velocity enables stable upstream operation against strong physiological flows representative of major arteries and veins, establishing the milli-spinner as a robust untethered navigation platform for operation in high-flow, tortuous vasculature.

## 1. Introduction

In recent decades, catheter- and guidewire-based interventional devices have become the primary tools for the diagnosis and treatment of life-threatening vascular diseases, including thrombosis, aneurysms, and atherosclerosis, which are highly prevalent among middle-aged and elderly populations [1-3]. For example, mechanical thrombectomy employs aspiration catheters and stent retrievers to remove intravascular clots [4, 5], while balloon catheters are widely used to dilate stenotic vessels [6, 7]. These devices are typically introduced via the femoral or radial artery and advanced to the target site through the vasculature. However, navigating such devices through highly tortuous pathways, such as cerebral arteries, remains extremely challenging, and many patients cannot be treated because the devices are unable to reach the disease site. In addition, excessive manipulation during catheter navigation poses a significant risk of vessel injury, including endothelial denudation and, in severe cases, vascular perforation.

Magnetically actuated milli-scale robots [8-13] offer a promising wireless strategy to overcome the limitations of catheter- and guidewire-based interventions by enabling controlled locomotion and guidance under externally applied magnetic fields. Among a variety of magnetic micro- and milli-robot designs for endovascular applications [12-17], propulsion in confined tubular environments is most commonly achieved using propeller-inspired helical structures [18-21] driven by rotating magnetic fields. Building on this concept, Wu et al. [22] recently introduced a magnetic milli-spinner featuring a cylindrical body with helical fins, a central through-hole, and side slits. As shown in **Figure 1a**, this unique architecture enables high propulsion velocities upon spinning, allowing the milli-spinner to swim against physiological blood flow with a controllable velocity by adjusting the rotating magnetic field frequency. For downstream navigation to perform therapeutic tasks, the milli-spinner's propulsion velocity $v_s$ is set slightly below the local flow velocity $v_f$. In this regime, the milli-spinner remains dynamically stable while being carried slowly downstream in a controlled manner by the flow, allowing precise positioning at the target site (e.g., a clot). After completing the therapeutic task, the milli-spinner retrieval is achieved by increasing the rotation frequency to raise the propulsion velocity above the flow velocity ($v_s > v_f$). This enables the milli-spinner to actively swim upstream, overcome the physiological flow, and re-enter the introducing sheath for safe retrieval. By tuning the rotation axis and rotating frequency of the applied magnetic field in real time, the milli-spinner achieves rapid, stable, precise, and fully wireless navigation in highly tortuous vascular networks, making it a versatile platform for multifunctional endovascular procedures, including efficient clot debulking, targeted drug delivery, and in situ aneurysm treatment (**Figure 1b**). Meanwhile, the central through-hole allows blood to pass through the milli-spinner, thereby avoiding vessel occlusion. In addition, the incorporation of the central through-hole and side slits generates a large pressure drop within the milli-spinner during rotation. This pressure drop plays a critical role in enabling high-velocity propulsion against physiological blood flow by reducing hydrodynamic resistance and enhancing propulsion efficiency. The resulting pressure gradient also produces a highly localized suction field that firmly draws thrombus onto the milli-spinner surface, enabling effective clot debulking. Under high-frequency rotation, the combined compression and shear forces densify the fibrin network, leading to clot shrinkage and facilitating efficient removal [22-24].

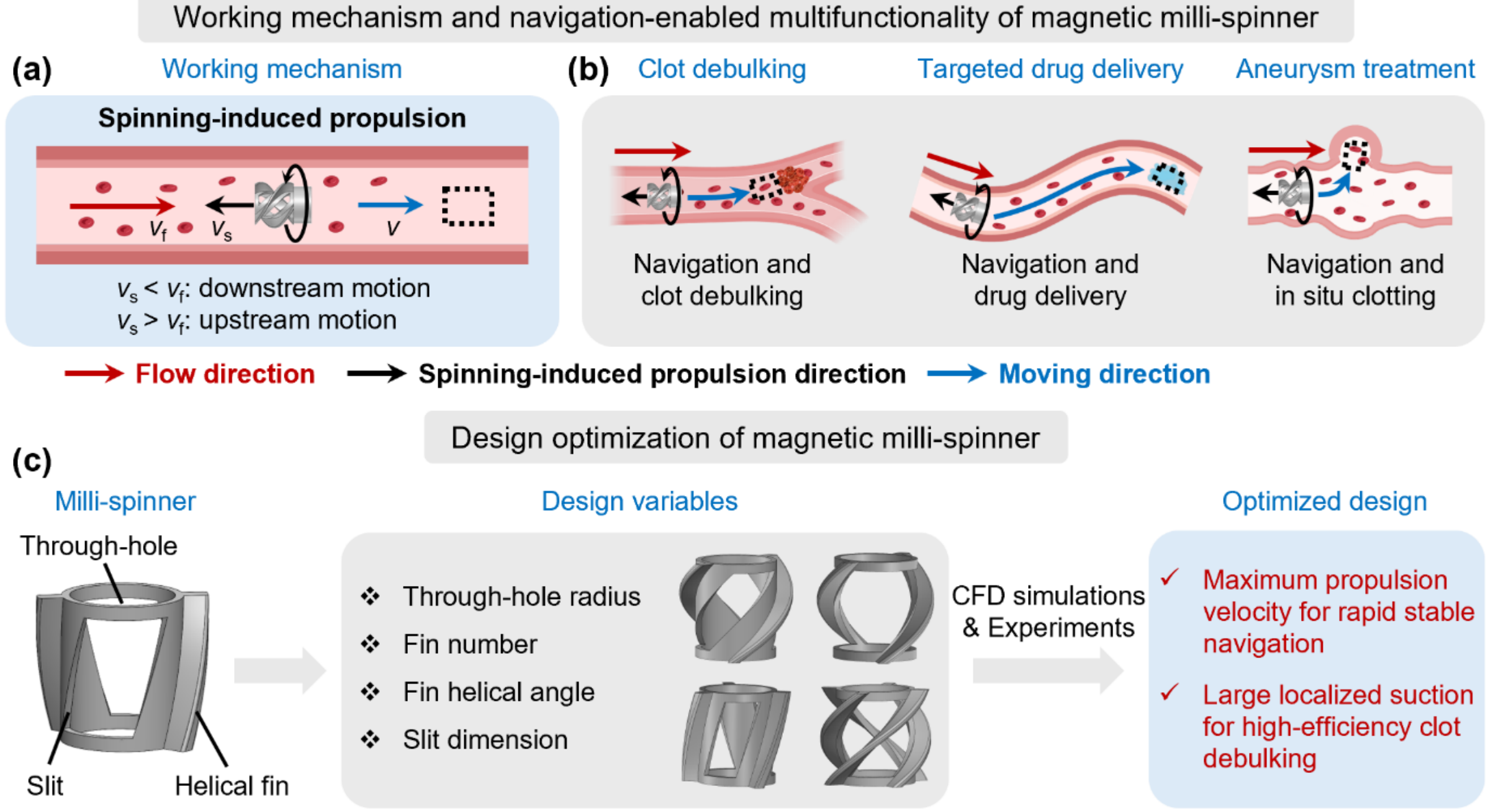

**Figure 1.** Multifunctional magnetic milli-spinner. (a) Working mechanism and (b) navigation-enabled multifunctionality of the magnetic milli-spinner in complex vascular environments. The propulsion direction of the milli-spinner is set against the blood flow direction to enable controllable navigation, and the absolute swimming velocity is regulated by tuning the rotating frequency of the applied magnetic field. For downstream navigation to perform therapeutic tasks, the spinning-induced propulsion velocity $v_s$ is set below the flow velocity $v_f$, such that the milli-spinner is carried downstream by the flow. For upstream swimming to remove the milli-spinner from the vessel by a sheath, $v_s$ is set beyond $v_f$, enabling upstream motion toward the device for retrieval. (c) Design optimization of the magnetic milli-spinner.

The high propulsion efficiency and clot debulking performance of the magnetic milli-spinner arise from the synergistic effect of three structural features: the through-hole, the slit, and the helical fin (**Figure 1c**). These elements introduce a rich set of tunable design parameters, which provide a vast design space for programming the magnetic milli-spinner's overall performance. Systematic parametric optimization of these parameters is therefore essential to fully exploit this potential. In this work, we aim to maximize the propulsion velocity of the magnetic milli-spinner for rapid, stable navigation in complex vascular environments with high flow rates, while simultaneously enhancing its internal pressure drop to generate large localized suction for high-efficiency clot debulking. To this end, we combine computational fluid dynamics (CFD) simulations with experimental validation to systematically investigate the effects of key geometric parameters, including through-hole radius, fin number, fin helical angle, and slit dimension, on propulsion efficiency and internal pressure drop. Guided by this parametric study, we identify an optimized magnetic milli-spinner design that achieves swimming velocities of 55 cm/s (~175 body lengths/s) in saline water and 44 cm/s (~140 body lengths/s) in a fluid with a viscosity of 3.5 mPa·s (comparable to that of arterial blood at high shear rates [25]), far beyond those of existing untethered magnetic robots in tubular environments (< 80 body lengths/s) [18-22]. Swimming performance of the optimized milli-spinner is further evaluated under different flow conditions (e.g., flow velocity and fluid viscosity) and different flow environments (e.g., steady flow and pulsatile flow). In addition, clot debulking capability of the optimized milli-spinner is demonstrated as a representative clinically relevant application. The results show that the milli-spinner can stably swim upstream against strong physiological flows representative of major arteries and veins (e.g., peak flow velocity of ~60 cm/s and average flow velocity of ~20-30 cm/s in the internal carotid artery [26], and peak flow velocity of ~40 cm/s and average flow velocity of ~10-20 cm/s in the inferior vena cava [27], which reduce by 10-30% due to anesthesia-induced decreases in cardiac output and blood pressure during surgical or interventional procedures [28]). These findings establish the magnetic milli-spinner as a viable platform for wireless endovascular intervention under physiologically relevant high-flow conditions and within highly tortuous vasculature, highlighting its potential as a versatile robotic system for a broad range of endovascular procedures, including robotic mechanical thrombectomy, embolectomy, and targeted drug delivery.

## 2. Results

As shown in **Figure 1c**, the geometry of the magnetic milli-spinner is defined by a rich set of parameters, including the outer diameter (OD), length, through-hole radius, wall thickness, fin length, fin thickness, fin helical angle, fin number, and slit dimension. Here, we focus on optimizing four key geometric parameters that dominate the propulsion performance: *the through-hole radius*, *fin number*, *fin helical angle*, and *slit dimension*. The remaining parameters are selected based on the fabrication constraints and the intended operating environment. The OD of the milli-spinner is fixed at 2.5 mm to enable operation in vessels such as coronary and cerebral arteries with diameters of 3-5 mm [29, 30]. The length of the milli-spinner is set to 2.15 mm, consistent with our original design [22] for comparison purposes. These two dimensions ensure proper fit and agile steering when navigating vessel bifurcations. The wall thickness and the fin thickness of the milli-spinner are set to 0.12 mm and 0.25 mm, respectively, based on the resolution limits of the 3D printing process. A detailed dimensional illustration of the milli-spinner design is provided in **Figure S1** in the Supporting Information. Notably, the milli-spinner geometry can be readily scaled up or down to accommodate a wide range of vessel sizes.

CFD simulations combined with experimental validation (see **Materials and methods** for details) are performed to systematically evaluate the propulsion performance of different milli-spinner designs. In the design optimization process (**Sections 2.1-2.3**), no background flow is imposed in the experiments, and the milli-spinner is placed inside a long tube with an inner diameter (ID) of 3.5 mm and an OD of 6.0 mm, filled with saline water and driven by a rotating magnetic field with a controllable frequency to achieve swimming propulsion. This tube inner diameter is comparable to that of medium-sized arteries, such as the middle cerebral artery (2-4 mm) [31]. In the experiments, the body of the magnetic milli-spinner is fabricated using a commercial 3D-printing Resin (Anycubic ABS-like Resin 3.0), and a ring-shaped neodymium magnet (grade N50, SM Magnetics, USA), with an ID of 1 mm, an OD of 2 mm, and a length of 1 mm, is coaxially attached to one end of the milli-spinner using adhesive for magnetic actuation. This ring magnet provides a high magnetization to the milli-spinner, and its interaction with the externally applied rotating magnetic field generates a sufficiently large magnetic torque, even under relatively low field strengths (e.g., 10-20 mT), to overcome the hydrodynamic resistance from the surrounding fluid and drive the milli-spinner to rotate synchronously with the applied magnetic field at the same frequency, thereby enabling effective and controllable propulsion in fluidic environments.

CFD simulations are conducted using COMSOL Multiphysics 6.1 (COMSOL Inc., USA). Zero-pressure boundary conditions are applied at both the inlet and outlet of the tube, as illustrated in **Figure S2** in the Supporting Information, and no-slip wall conditions are imposed on the milli-spinner surface and tube wall. Stationary analysis is performed to determine the steady-state flow field, the pressure profile, and the resulting forces acting on the milli-spinner. To determine the propulsion velocity, an initial velocity is applied to the tube wall and iteratively adjusted until the milli-spinner reaches equilibrium. In the equilibrium state, the axial resultant force (i.e., the stress integration in the longitudinal direction over the milli-spinner's surface) approaches zero, indicating that the prescribed wall velocity corresponds to the magnitude of the milli-spinner's equilibrium propulsion velocity but in the opposite direction. This velocity also represents the maximum constant flow velocity that the milli-spinner can overcome.

### 2.1. Effect of through-hole radius

As stated previously, the geometric feature combining a central through-hole and side slits significantly enhances propulsion efficiency and clot debulking capability by generating a pressure drop within the milli-spinner. Moreover, the through-hole serves as a flow channel that allows blood to pass through the milli-spinner, thereby preventing vessel occlusion. We therefore begin design optimization by examining the effect of the through-hole radius. As shown in **Figure 2a**, with the milli-spinner's OD fixed at 2.5 mm, the through-hole radius-to-fin length ratio, $R_{in}/L_{fin}$, is selected as the design parameter and varied from 0.5 to 4. Three fin numbers are considered, i.e., $N = 2$, 3, and 4. For all cases, the fin helical angle is fixed at 30°, which lies within the typical range (20-40°) used in propeller designs [32]. In addition, the slit number equals the fin number. Each slit has two long edges parallel to the helical fin and two short edges parallel to the cylinder ends, with a fixed distance of 1.75 mm between the two short edges (**Figure 2a**). The width of the short edge of a single slit is adjusted according to the fin number to maintain a constant total slit width of 1.83 mm (see details in **Section S1**, **Figure S1**, and **Tables S1-S3** in the Supporting Information).

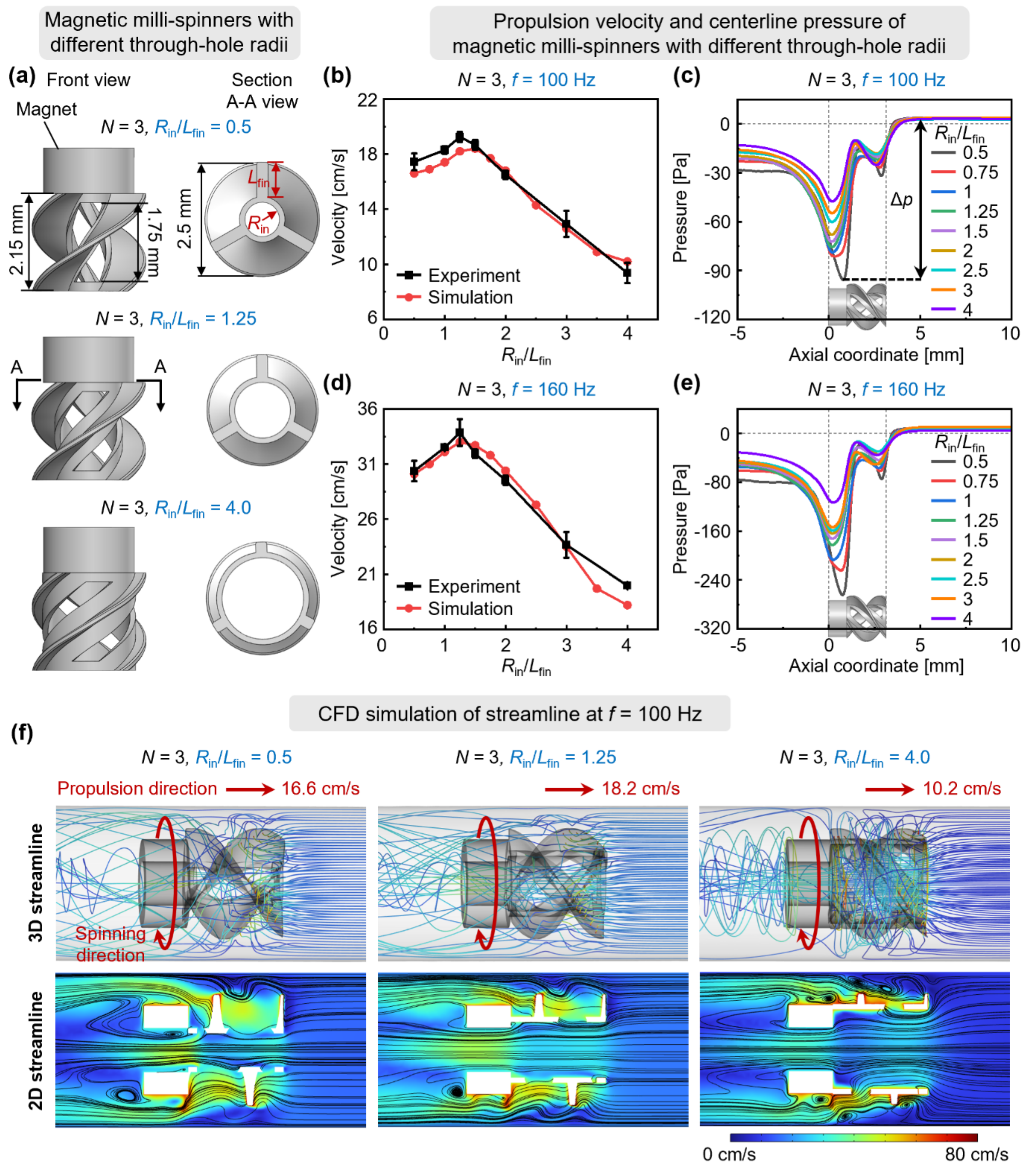

**Figure 2.** Effect of the through-hole radius on the propulsion performance of 3-fin magnetic milli-spinners in a 3.5 mm-ID tube. (a) Schematics of 3-fin milli-spinners with different through-hole radius-to-fin length ratios $R_{in}/L_{fin}$. (b) Propulsion velocity and (c) centerline pressure of 3-fin milli-spinners with different $R_{in}/L_{fin}$ under a rotating magnetic field with frequency $f$ = 100 Hz. (d) Propulsion velocity and (e) centerline pressure of 3-fin milli-spinners with different $R_{in}/L_{fin}$ under a rotating magnetic field with frequency $f$ = 160 Hz. (f) 3D streamlines starting from the tube inlet and 2D streamlines on the longitudinal cut plane for 3-fin milli-spinners with different $R_{in}/L_{fin}$ at $f$ = 100 Hz.

The effect of the through-hole radius on the propulsion performance of 3-fin magnetic milli-spinners in a 3.5 mm-ID tube under a rotating magnetic field is investigated in **Figure 2** through CFD simulations and experimental validation. Two rotating-field frequencies, $f$ = 100 Hz and 160 Hz, are examined for actuation. As shown in **Figures 2b** and **2d,** at both frequencies, the simulations predict that the propulsion velocity of the milli-spinner first increases and then decreases as the through-hole radius increases, reaching maximum values of 18.4 cm/s and 33.0 cm/s at $R_{in}/L_{fin}$ = 1.5 and 1.25 for $f$ = 100 Hz and 160 Hz, respectively. The experimental results show a similar nonmonotonic trend, with maximum velocities of 19.3 ± 0.35 cm/s and 33.9 ± 1.21 cm/s at $f$ = 100 Hz and 160 Hz, respectively, when $R_{in}/L_{fin}$ = 1.25. In particular, the velocity at 160 Hz is 47.4% higher than the maximum velocity of 23.0 cm/s achieved by our original design with the same OD and length [22], highlighting the critical role of design optimization in enhancing the milli-spinner's propulsion efficiency. Experimental demonstrations of the propulsion performance of 3-fin magnetic milli-spinners with $R_{in}/L_{fin}$ = 0.5, 1.25, and 4.0 are shown in **Figure 3** and **Movie S1** (Supporting Information). The measured velocities of the three designs at $f$ = 160 Hz are 30.8 cm/s, 35.0 cm/s, and 20.1 cm/s, respectively. As expected, the milli-spinner with $R_{in}/L_{fin}$ = 1.25 travels a longer distance than the other two designs within the same time interval. Combining the simulation and experimental results, the optimal through-hole radius for the 3-fin milli-spinner to achieve the maximum propulsion velocity is $R_{in}/L_{fin}$ = 1.25.

The centerline pressure along the milli-spinner at the two frequencies predicted by CFD simulations is shown in **Figures 2c** and **2e**. In all cases, a pressure drop $\Delta p$ (i.e., the difference between the pressure at the distal end of the milli-spinner front and the lowest pressure within the milli-spinner) is observed inside the milli-spinner, resulting from the combined effects of the through-hole and the slits. At both frequencies, $\Delta p$ decreases monotonically as $R_{in}/L_{fin}$ increases from 0.5 to 4, indicating that a smaller through-hole combined with longer fins produces a higher pressure drop and thereby generates larger localized suction for high-efficiency clot debulking. The optimal through-hole radius for maximum pressure drop differs from that for maximum propulsion velocity. Since the primary focus of this study is to improve the propulsion efficiency of the milli-spinner, the optimal through-hole radius for 3-fin magnetic milli-spinners is selected as $R_{in}/L_{fin}$ = 1.25.

It should be noted that the ring magnet used to actuate the milli-spinner has a significant influence on both the propulsion velocity and centerline pressure. **Figure S3** in the Supporting Information presents the propulsion velocity and centerline pressure of 3-fin milli-spinners without the ring magnet for different through-hole radii based on CFD simulations. The results show that the ring magnet decreases the propulsion velocity but increases the pressure drop of the milli-spinner. This suggests that the propulsion efficiency of the milli-spinner could be further improved by optimizing the actuation strategy, such as distributing multiple smaller magnets at both ends of the milli-spinner [22].

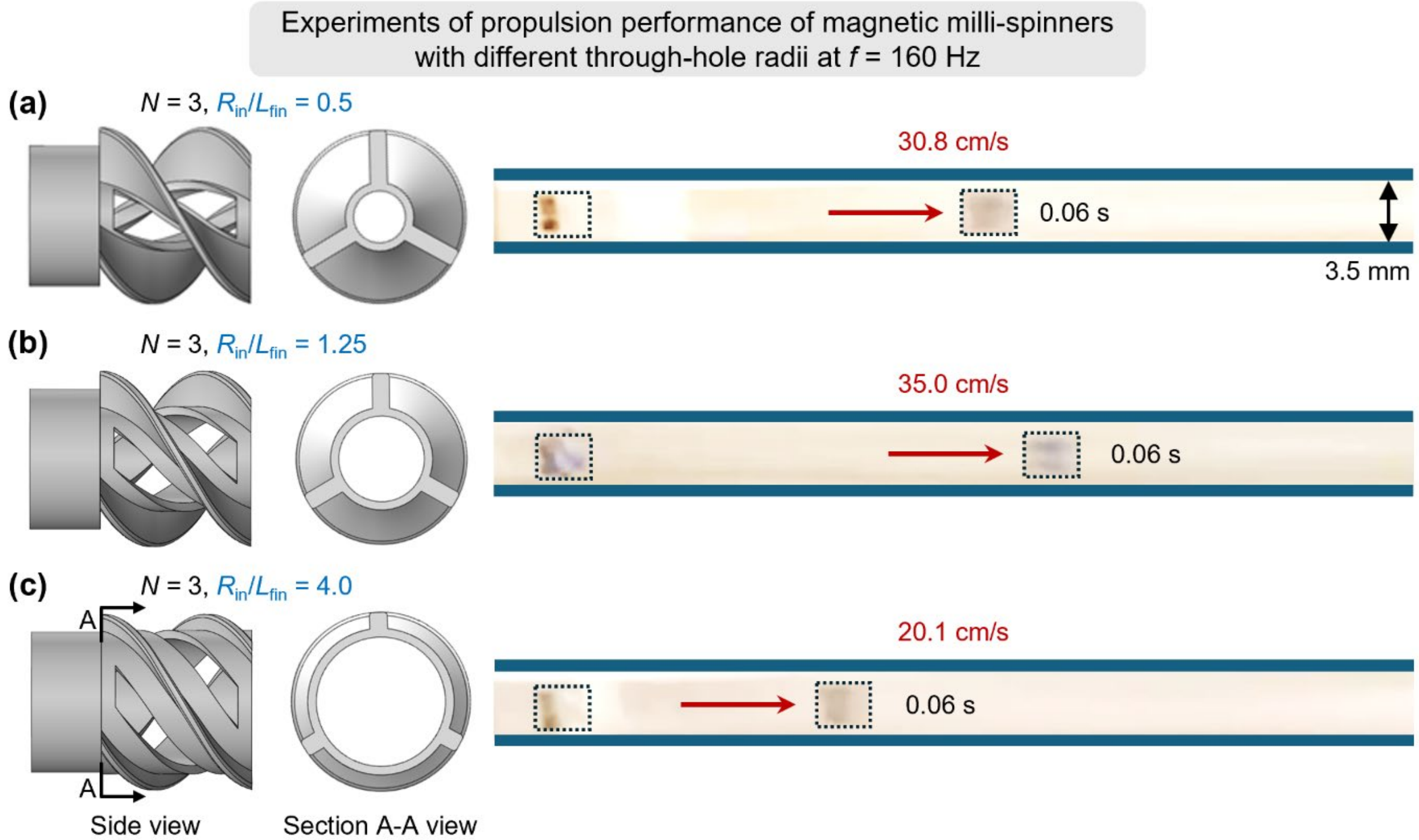

**Figure 3.** Experiments of propulsion performance of 3-fin magnetic milli-spinners with different through-hole radii in a 3.5 mm-ID tube under a 160 Hz rotating magnetic field. (a) $R_{in}/L_{fin} = 0.5$. (b) $R_{in}/L_{fin} = 1.25$. (c) $R_{in}/L_{fin} = 4$. Experimental results demonstrate that the 3-fin milli-spinner achieves the highest propulsion velocity when the through-hole radius-to-fin length ratio is $R_{in}/L_{fin} = 1.25$.

To elucidate the distinct effects of $R_{in}/L_{fin}$ on the propulsion efficiency of magnetic milli-spinners, **Figure 2f** compares the CFD-predicted 2D and 3D streamlines in the tube for different through-hole radii at $f = 100$ Hz. For small or large through-hole radii (e.g., $R_{in}/L_{fin} = 0.5$ and 4), strong vortices are observed near the tube wall and at the rear of the milli-spinner during spinning motion, as evident in the 2D streamlines, which dissipate energy and reduce propulsion efficiency. In contrast, with an intermediate through-hole radius (e.g., $R_{in}/L_{fin} = 1.25$), the spinning motion generates more coherent axial streamlines, leading to a continuous high-velocity region in the tube, as shown in the 3D streamlines.

Results for 2-fin and 4-fin magnetic milli-spinners with different through-hole radii are provided in **Figures S4** and **S5** in the Supporting Information. The 2-fin milli-spinner achieves its maximum velocities of 18.8 ± 1.0 cm/s and 32.8 ± 0.46 cm/s at $f$=100 Hz and 160 Hz, respectively, when $R_{in}/L_{fin}$ = 0.5. The 4-fin milli-spinner reaches its maximum velocities of 14.3 ± 0.20 cm/s and 27.0 ± 0.68 cm/s at $f$ = 100 Hz and 160 Hz, respectively, when $R_{in}/L_{fin}$ = 1.5. Compared with the 3-fin milli-spinner at $R_{in}/L_{fin}$ = 1.25 (19.3 ± 0.35 cm/s and 33.9 ± 1.21 cm/s at $f$ = 100 Hz and 160 Hz, respectively), the 2-fin and 4-fin designs exhibit lower velocities at both frequencies (see experimental demonstrations in **Movie S2**, Supporting Information). Based on these results, the 3-fin magnetic milli-spinner with $R_{in}/L_{fin}$ = 1.25 is selected as the optimized design and is used for subsequent optimizations of the fin helical angle and slit dimension.

**2.2. Effect of fin helical angle**

Next, we optimize the fin helical angle based on the first-step optimization results of the through-hole radius. Specifically, the fin number and the through-hole radius-to-fin length ratio are fixed at $N$ = 3 and $R_{in}/L_{fin}$ = 1.25, respectively. The width of a single slit is kept constant at 0.61 mm for all cases. To identify the optimal fin helical angle, we compare the propulsion velocity and centerline pressure of 3-fin magnetic milli-spinners with helical angles ranging from 20° to 80° (**Figure 4a**) in a 3.5 mm-ID tube. **Figures 4b** and **4d** show the propulsion velocities of milli-spinners with different helical angles under rotating magnetic fields of 100 Hz and 160 Hz, respectively. In both cases, simulations and experiments consistently show that the propulsion velocity first increases and then decreases as the helical angle $\alpha$ increases from 20° to 80°, reaching its maximum at $\alpha$ = 60°. The maximum experimental velocities are 23.8 ± 0.53 cm/s at $f$ = 100 Hz and 45.0 ± 1.87 cm/s at $f$ = 160 Hz, which are enhanced by 23.3% and 32.7%, respectively, compared with the first-step optimized design. Experimental demonstrations of the propulsion performance of magnetic milli-spinners with $\alpha$ = 30°, 60°, and 80° are presented in **Figure 5** and **Movie S3** (Supporting Information). At $f$ = 160 Hz, the measured propulsion velocities are 35.0 cm/s, 45.6 cm/s, and 30.0 cm/s for $\alpha$ = 30°, 60°, and 80°, respectively. These results confirm that the optimal fin helical angle of 3-fin milli-spinners with $R_{in}/L_{fin}$ = 1.25 for achieving the highest propulsion efficiency is $\alpha$ = 60°.

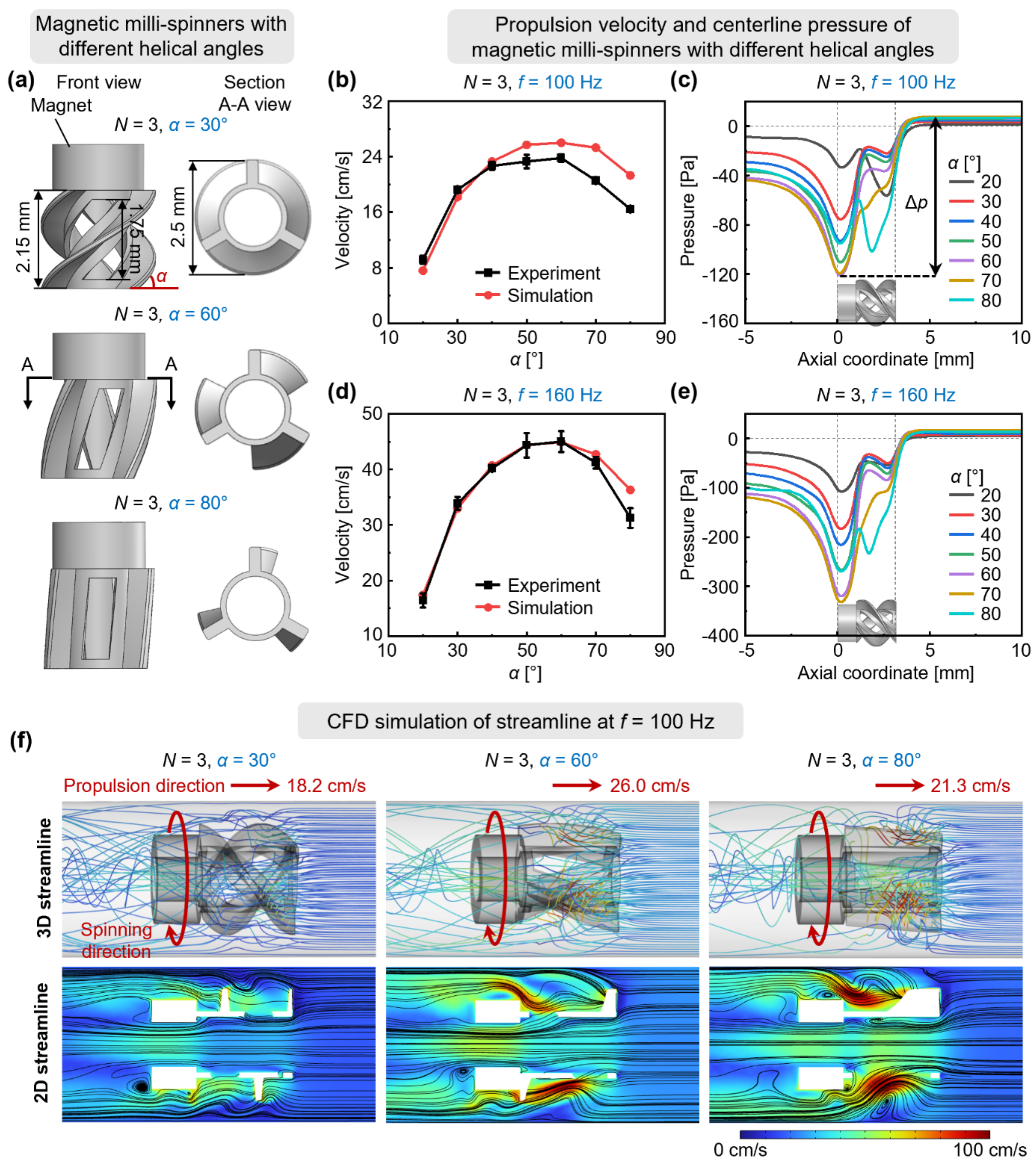

**Figure 4.** Effect of the fin helical angle on the propulsion performance of 3-fin magnetic milli-spinners with $R_{in}/L_{fin} = 1.25$ in a 3.5 mm-ID tube. (a) Schematics of 3-fin milli-spinners with different fin helical angles $\alpha$. (b) Propulsion velocity and (c) centerline pressure of 3-fin milli-spinners with different helical angles under a rotating magnetic field with frequency $f = 100$ Hz. (d) Propulsion velocity and (e) centerline pressure of 3-fin milli-spinners with different helical angles under a rotating magnetic field with frequency $f = 160$ Hz. (f) 3D streamlines starting from the tube inlet and 2D streamlines along the longitudinal cut plane for 3-fin milli-spinners with different helical angles at $f = 100$ Hz.

The centerline pressure along the milli-spinner inside the tube at the two rotating frequencies predicted by CFD simulations is shown in **Figures 4c** and **4e**. At both frequencies, the pressure drop $\Delta p$ varies nonmonotonically as $\alpha$ increases from 20° to 80°, with maximum $\Delta p$ occurring at $\alpha$ = 60° and 70° for $f$ = 100 Hz and 160 Hz, respectively. Compared with the first-step optimization of the through-hole radius, the maximum pressure drops increase by 25.0% and 25.3% at the two frequencies after optimizing the fin helical angle, resulting in stronger localized suction and enhanced clot-debulking capability. Notably, the helical angle for maximum propulsion velocity is nearly consistent with that for maximum pressure drop. Therefore, the optimal helical angle for 3-fin milli-spinners with $R_{in}/L_{fin}$ = 1.25 is selected as $\alpha$ = 60°.

The effects of the fin helical angle on the propulsion velocity and centerline pressure of 3-fin milli-spinners without a ring magnet are examined using CFD simulations in **Figure S6** in the Supporting Information. As expected, at the same rotating frequency, the milli-spinner without a ring magnet exhibits higher propulsion velocity than that with a ring magnet, owing to reduced weight and fluid resistance. In the absence of a ring magnet, the variation trends of both propulsion velocity and pressure drop with respect to the helical angle are similar to those observed with a ring magnet. However, the helical angles corresponding to the maximum propulsion velocity and maximum pressure drop shift to $\alpha$ = 50° and $\alpha$ = 70°, respectively, at both frequencies.

**Figure 4f** illustrates the CFD-predicted 3D and 2D streamlines inside the tube for different helical angles at $f$ = 100 Hz. From the 3D streamlines, it is observed that when the helical angle is relatively small (e.g., $\alpha$ = 30°), the spinning-induced flow velocity is noticeably lower than that observed for $\alpha$ = 60° and 80°. In these two cases, the flow passing through the central through-hole and side slits is significantly accelerated, as indicated by the red regions in the 2D streamlines, creating a larger pressure drop that reduces the hydrodynamic resistance and improves the propulsion efficiency. In addition, for $\alpha$ = 80°, stronger vortices form near the tube wall. These vortices dissipate energy and decrease propulsion efficiency. As a result, the milli-spinner with $\alpha$ = 60° achieves the highest propulsion velocity among the three cases.

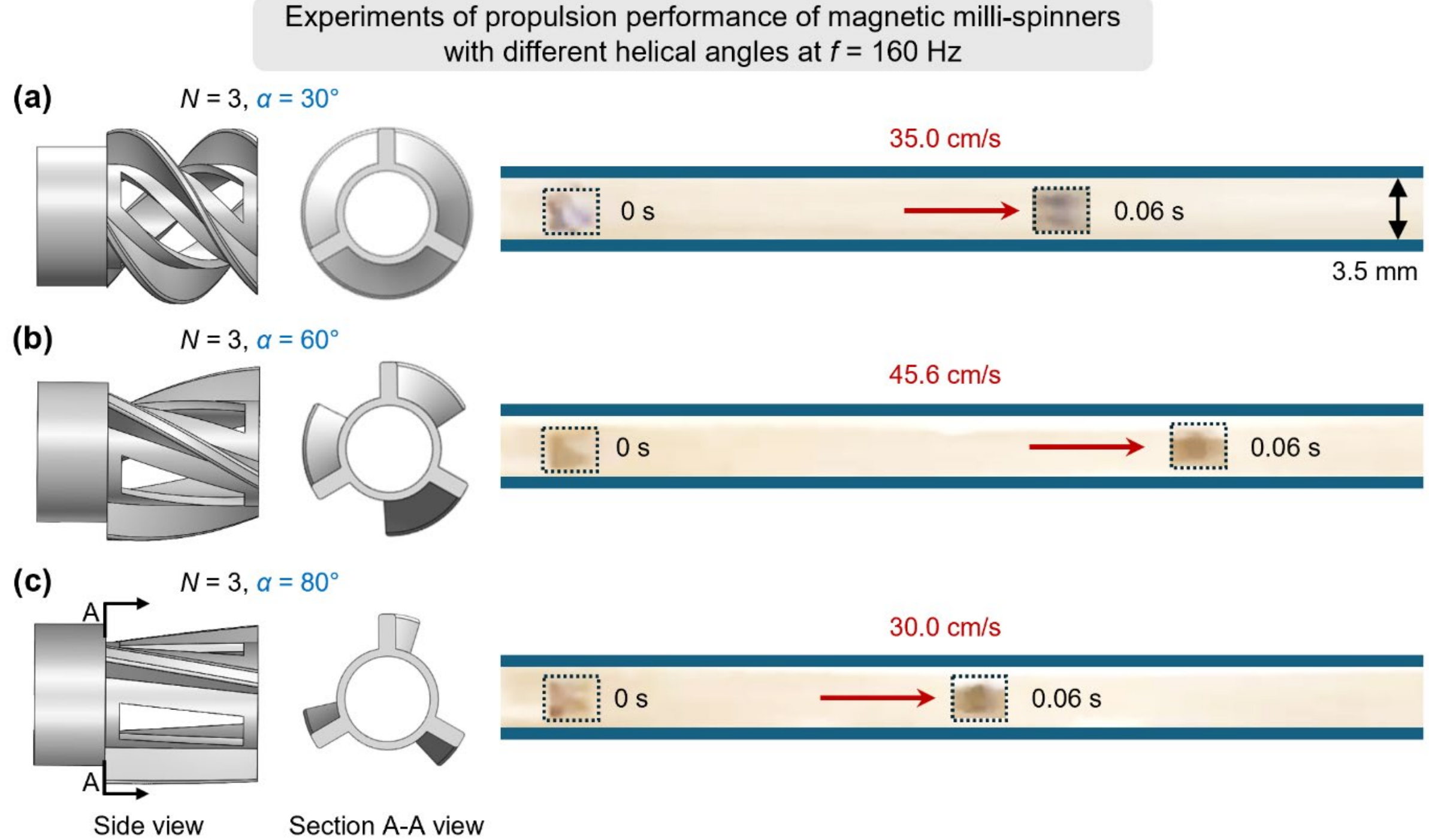

**Figure 5.** Experiments of propulsion performance of 3-fin magnetic milli-spinners with $R_{in}/L_{fin} = 1.25$ and different fin helical angles in a 3.5 mm-ID tube under a 160 Hz rotating magnetic field. (a) $\alpha = 30°$. (b) $\alpha = 60°$. (c) $\alpha = 80°$. Experimental results demonstrate that the 3-fin milli-spinner with $R_{in}/L_{fin} = 1.25$ achieves the highest propulsion velocity when the helical angle is $\alpha = 60°$.

## 2.3. Effect of slit dimension

In the following, we further optimize the slit dimension by examining its effects on the propulsion velocity and centerline pressure of magnetic milli-spinners. Based on the first two-step optimization, the fin number, the through-hole radius-to-fin length ratio, and the fin helical angle are fixed at $N = 3$, $R_{in}/L_{fin} = 1.25$, and $\alpha = 60°$, respectively. As shown in **Figure 6a**, consistent with the slit design used in the previous optimization steps, each slit has two long edges parallel to the helical fin and two short edges parallel to the cylinder ends, with a fixed distance of 1.75 mm between the two short edges. The total width $w_T$ of the short edges of the three slits is normalized by the cylinder centerline circumference $S = 2\pi r$, where $r$ is the centerline radius. This normalized width $w_T/S$ is set as the design variable and varied from 0.25 to 0.875 (see details in **Table S4** in the Supporting Information, and $w_T/S = 0.875$ corresponds to the maximum slit width that can be designed for 3-fin milli-spinners with a fin thickness of 0.25 mm).

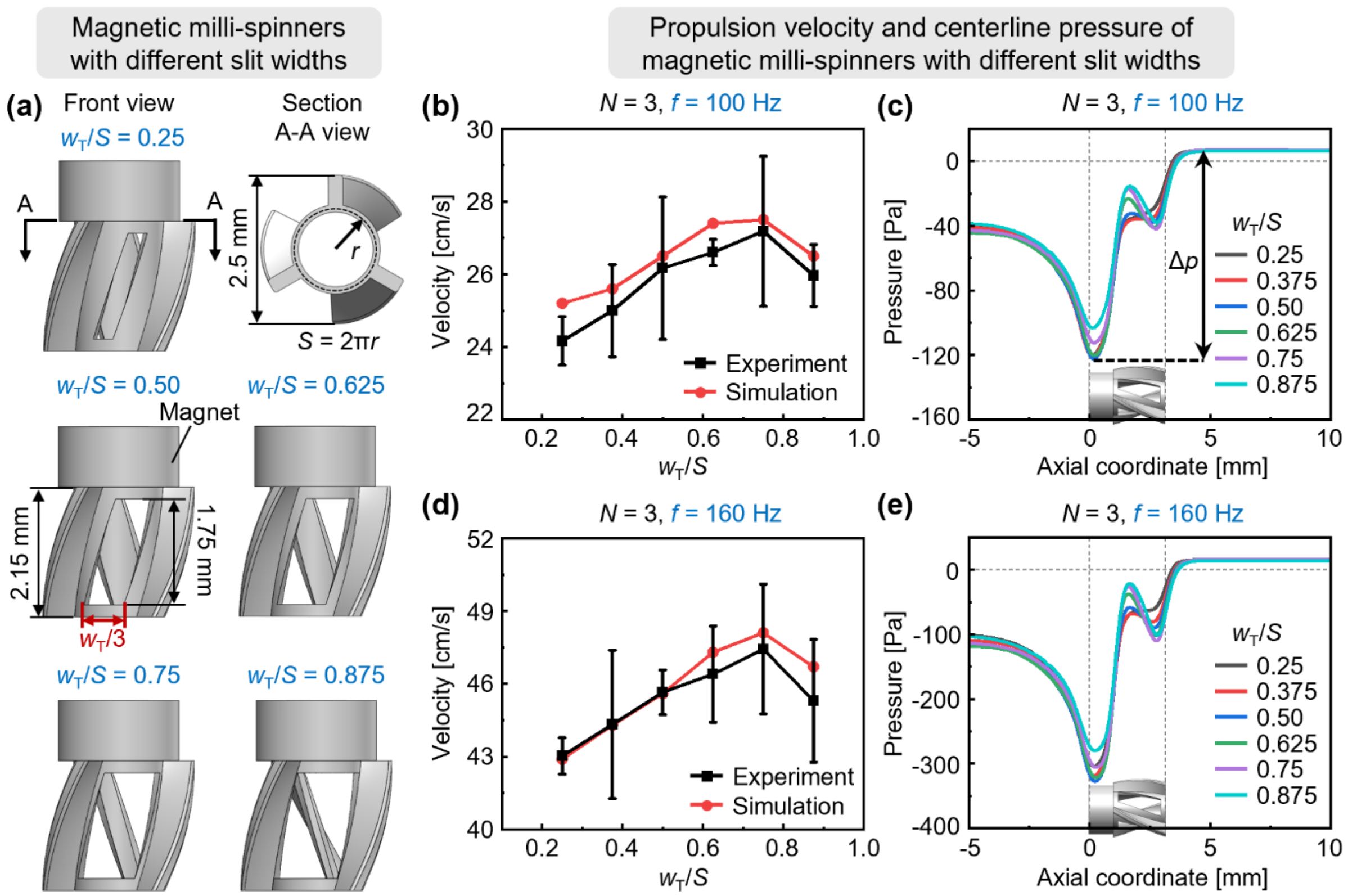

**Figure 6.** Effect of the slit width on the propulsion performance of 3-fin magnetic milli-spinners in a 3.5 mm-ID tube. (a) Schematics of 3-fin milli-spinners with varying normalized slit width $w_T/S$. (b) Propulsion velocity and (c) centerline pressure of 3-fin milli-spinners with different $w_T/S$ under a rotating magnetic field with frequency $f = 100$ Hz. (d) Propulsion velocity and (e) centerline pressure of 3-fin milli-spinners with different $w_T/S$ under a rotating magnetic field with frequency $f = 160$ Hz.

**Figures 6b** and **6d** show the propulsion velocities of magnetic milli-spinners with different slit widths in a 3.5 mm-ID tube under rotating magnetic fields of 100 Hz and 160 Hz, respectively. After the first two-step optimization of through-hole radius and fin helical angle, the effect of slit dimension on propulsion efficiency becomes less significant. Both simulations and experiments indicate that the milli-spinner achieves its maximum velocity when the normalized slit width $w_T/S$ increases to 0.75 at the two frequencies. The experimental maximum velocities at $f = 100$ Hz and 160 Hz are 27.2 ± 2.06 cm/s and 47.4 ± 2.68 cm/s, respectively, which are enhanced by 14.3% and 5.3% compared with the velocities obtained after optimizing the helical angle. Experimental demonstrations of the propulsion performance of milli-spinners with $w_T/S = 0.25$, 0.75, and 0.875 are shown in **Figure 7** and **Movie S4** (Supporting Information). At $f = 160$ Hz, the measured propulsion velocities for milli-spinners with $w_T/S = 0.25$, 0.75, and 0.875 are 43.0 cm/s, 50.3 cm/s, and 44.9 cm/s, respectively. Combining simulation and experimental results, the optimal normalized slit width of 3-fin milli-spinners with $R_{in}/L_{fin}$ = 1.25 and $\alpha$ = 60° for achieving the maximum propulsion velocity is $w_T/S$ = 0.75.

The corresponding centerline pressure along the milli-spinner inside the tube for different slit widths at $f$ = 100 Hz and 160 Hz are shown in **Figures 6c** and **6e**, respectively. Similar to the propulsion velocity, the pressure drop exhibits a relatively weak dependence on slit width after optimization of the through-hole radius and fin helical angle. For both frequencies, the pressure drop reaches its maximum at $w_T/S$ = 0.5. As the primary objective is to maximize the propulsion efficiency, the optimal normalized slit width of 3-fin milli-spinners with $R_{in}/L_{fin}$ = 1.25 and $\alpha$ = 60° is therefore selected as $w_T/S$ = 0.75, which is also the final optimized design after the three-step optimization. It should be noted that, although the slit width has a relatively minor effect on the propulsion velocity and internal pressure drop of the milli-spinner with optimized through-hole radius and fin helical angle, a larger slit width contributes to a higher drug delivery rate by providing a larger area for fluid exchange between the drug within the internal cavity and the surrounding environment [22]. Therefore, for applications involving drug delivery, a larger slit width may be considered in the milli-spinner design.

The effects of slit dimension on the propulsion velocity and centerline pressure of milli-spinners without a ring magnet are studied in **Figure S7** in the Supporting Information using CFD simulations. In this case, the propulsion velocity also becomes less sensitive to the slit width after the first two-step optimization, but the velocity is increased by approximately 20% compared to milli-spinners with a ring magnet. However, in the absence of a ring magnet, the pressure drop decreases monotonically with increasing slit width, suggesting that narrower slits lead to a higher pressure drop, which generates larger localized suction that facilitates clot debulking.

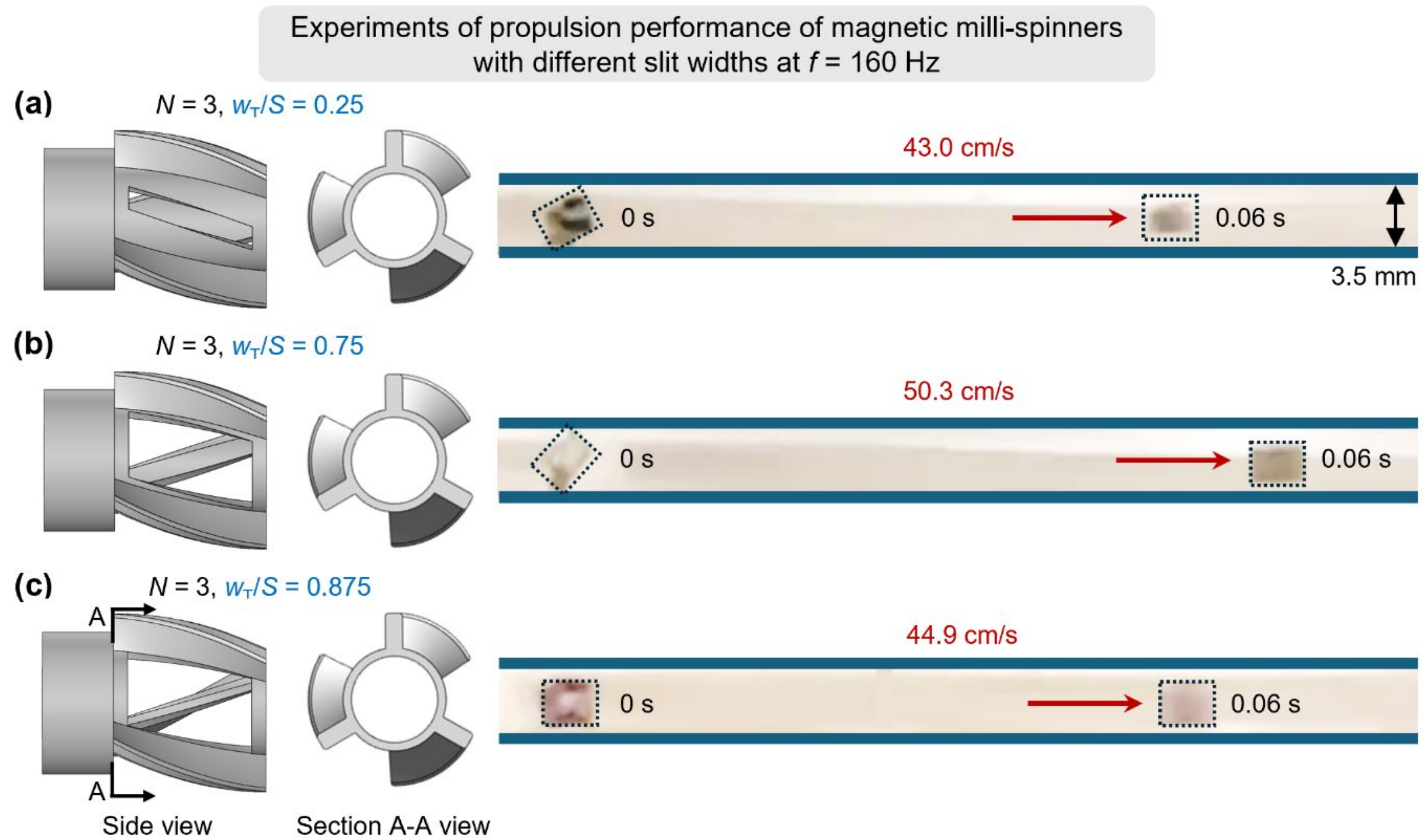

**Figure 7.** Experiments of propulsion performance of 3-fin magnetic milli-spinners with different slit widths in a 3.5 mm-ID tube under a 160 Hz rotating magnetic field. (a) $w_T/S$ = 0.25. (b) $w_T/S$ = 0.75. (c) $w_T/S$ = 0.875. Experimental results demonstrate that the 3-fin milli-spinner with $R_{in}/L_{fin}$ = 1.25 and $\alpha$ = 60° achieves the highest propulsion velocity when the normalized slit width is $w_T/S$ = 0.75.

## 2.4. Optimized magnetic milli-spinner for high-efficiency propulsion and clot debulking

Next, we evaluate the propulsion performance of the optimized magnetic milli-spinner (**Figure 8a**) in fluids with varying viscosities and at different rotating frequencies. To examine the effect of fluid viscosity on propulsion performance, the optimized milli-spinner is tested in a 3.5 mm-ID tube filled with a saline water-glycerin mixture in the absence of background flow. As mentioned earlier, this tube inner diameter is comparable to that of medium-sized arteries, such as the middle cerebral artery. In the experiments, the volume fraction of glycerin is varied from 0% to 10%, 20%, and 30% to change the fluid viscosity, resulting in viscosities of 1.0, 1.5, 2.5, and 3.5 mPa·s [23], respectively. As shown in **Figure 8b**, the propulsion velocity of the milli-spinner gradually decreases with increasing fluid viscosity due to increased hydrodynamic resistance. At a fluid viscosity of 3.5 mPa·s, corresponding to a mixture of 70 vol.% saline water and 30 vol.% glycerin and falling within the range of 3-4 mPa·s for arterial blood at high shear rates [25], the propulsion velocity decreases by approximately 30% compared to that in saline water (1.0 mPa·s).

The propulsion velocities of the milli-spinner at different rotating frequencies in saline water (1.0 mPa·s) and in the saline water-glycerin mixture with a viscosity of 3.5 mPa·s, without background flow, are shown in **Figure 8c**. In both fluids, simulations and experiments consistently show that the propulsion velocity (as well as the internal pressure drop, see **Figure S8** in the Supporting Information) increases nearly linearly with the rotating frequency, indicating that higher actuation frequencies enhance both propulsion efficiency and clot debulking capability. At $f$ = 180 Hz (the highest frequency achievable with our coil system), the maximum experimental velocity reaches 55 cm/s in saline water (~175 body lengths/s, see **Figure 8d** and **Movie S5** in the Supporting Information) and 44 cm/s (~140 body lengths/s, see **Movie S5**) in a saline water-glycerin mixture with a viscosity of 3.5 mPa·s (comparable to that of arterial blood at high shear rates [25]), far beyond those of existing untethered magnetic robots in tubular environments (< 80 body lengths/s) [18-22]. Note that even higher propulsion velocities could be achieved by further increasing the rotating frequency (**Figure 8c**). Moreover, the magnetic milli-spinner design is inherently scalable. Through structural optimization of its key geometric features, the optimized design can be directly miniaturized or scaled up to accommodate different vessel sizes and application requirements while maintaining effective propulsion performance.

As a representative clinically relevant application, the clot debulking capability of the optimized magnetic milli-spinner for robotic mechanical thrombectomy is demonstrated in **Figure 8e** and **Movie S6** (Supporting Information). The clot debulking experiment is conducted in a 3.5 mm-ID tube filled with saline water without background flow, and the milli-spinner is actuated by a 160 Hz rotating magnetic field generated by a manually controlled, motor-driven rotating magnet. As shown in **Figure 8e** and **Movie S6**, the milli-spinner first propels itself to the whole blood clot, after which the clot is drawn against the milli-spinner due to the suction force generated by the internal pressure drop. Simultaneously, the milli-spinner's rotation applies a continuous shear force to the clot. The combined compression and shear forces result in drastic densification and collapse of the fibrin network while expelling red blood cells (RBCs) [22-24], reducing the clot volume by approximately 50% within 40 s. Meanwhile, the strong suction force draws the entire clot proximally during the debulking process, effectively anchoring it and preventing distal migration or washout by physiological blood flow under realistic operating conditions. After treatment for 67 s, the clot volume is ultimately reduced by approximately 97%, and the clot transforms into a highly densified white fibrin core (**Figure 8f**), demonstrating the high-efficiency clot debulking capability of the optimized magnetic milli-spinner.

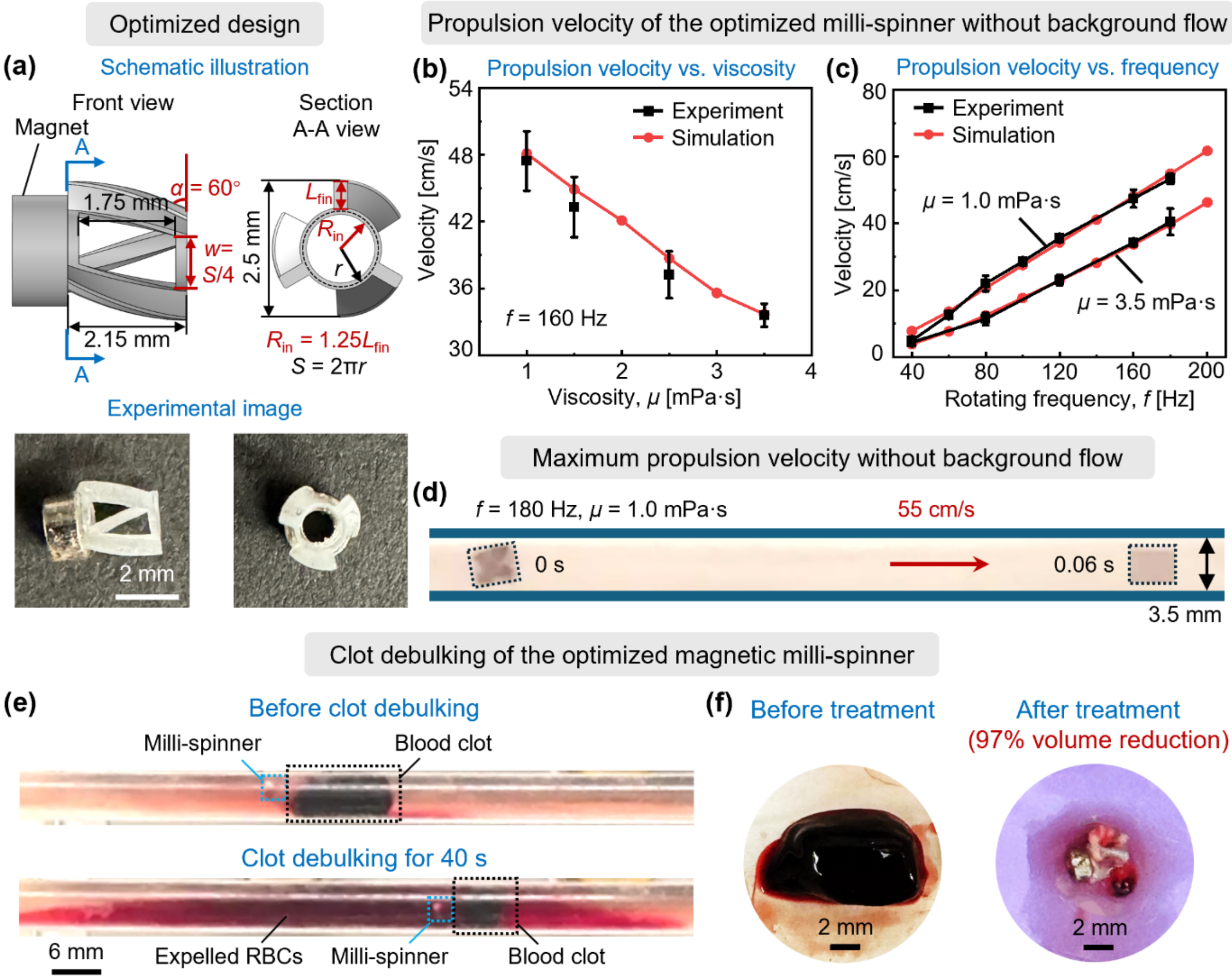

**Figure 8.** Propulsion performance and clot debulking of the optimized magnetic milli-spinner in a 3.5 mm-ID tube. (a) Schematic illustration and experimental image of the optimized milli-spinner design. (b) Effect of fluid viscosity on the propulsion velocity of the milli-spinner at a rotating magnetic field frequency of $f$ = 160 Hz without background flow. (c) Propulsion velocity of the milli-spinner as a function of the rotating magnetic field frequency in saline water or in a saline water-glycerin mixture with a viscosity of 3.5 mPa·s. (d) Experimental demonstration of the maximum propulsion velocity of the milli-spinner in saline water without background flow. (e) Clot debulking of the optimized magnetic milli-spinner without background flow. During the debulking process, the fibrin network of the clot undergoes significant densification and collapse, with RBCs being expelled. (f) Experimental images of the whole blood clot before and after treatment. After treatment, the clot transforms into a highly densified white fibrin core and exhibits an approximately 97% reduction in volume.

## 2.5. Swimming performance of the optimized magnetic milli-spinner under steady and pulsatile flows

The exceptional propulsion velocity of the optimized magnetic milli-spinner enables stable swimming in vascular environments with strong physiological flows representative of major arteries and veins. For instance, in the internal carotid artery, the peak blood flow velocity can reach ~60 cm/s, with an average velocity of ~20-30 cm/s [26]. In the inferior vena cava, the peak and average flow velocities are ~40 cm/s and ~10-20 cm/s, respectively [27]. During surgical or interventional procedures, these blood flow velocities are often reduced by 10-30% due to anesthesia-induced decreases in cardiac output and blood pressure [28]. To validate the swimming capability of the milli-spinner under clinically relevant conditions, in this section, its performance is experimentally evaluated in a 3.5 mm-ID tube under both steady and pulsatile flows. In these experiments, a saline water-glycerin mixture (70 vol.% saline water and 30 vol.% glycerin) with a viscosity of 3.5 mPa·s, comparable to that of arterial blood at high shear rates [25], is used to mimic physiological fluid properties.

**Figure 9a** and **9b** present the upstream propulsion performance of the milli-spinner at $f$ = 180 Hz under steady flows with flow velocities ranging from 20 to 30 cm/s. As the imposed flow velocity increases, the absolute upstream swimming velocity of the milli-spinner decreases (**Figure 9a** and **Movie S7** in the Supporting Information). Under flow velocities of 20 cm/s and 25 cm/s, the milli-spinner successfully overcomes the imposed flow and achieves absolute upstream swimming velocities of 12.8 cm/s and 5.7 cm/s (**Figure 9b**), respectively. These results indicate that the optimized magnetic milli-spinner is capable of stable upstream propulsion in steady vascular flow environments characterized by high flow velocities and physiologically relevant viscosities.

The upstream propulsion performance of the milli-spinner under a 60 beats/min pulsatile flow, with a peak velocity of 34 cm/s and an average velocity of 17 cm/s, is demonstrated in **Figure 9c** and **Movie S7**. In these two experiments, the milli-spinner achieves prescribed average upstream swimming velocities ($v_{avg}$) of 2 cm/s and 5 cm/s, respectively, by rationally selecting the actuation frequency of the rotating magnetic field. Such low target velocities are typically preferred during endovascular interventions to ensure procedural safety, full controllability, and precise navigation. **Figure 9c** shows the corresponding displacement-time curves, along with linear fits used to extract the average upstream swimming velocity. Owing to the periodic acceleration and deceleration of the pulsatile flow, the milli-spinner undergoes forward and backward motion within each pulsatile cycle, resulting in net upstream propulsion over time. By setting the rotating magnetic field frequency to 100 Hz and 110 Hz, the milli-spinner attains average upstream swimming velocities of $v_{avg}$ = 2.3 cm/s and 5.4 cm/s, respectively, showing good agreement with the prescribed target values. Such controllable low-velocity motion under pulsatile flow provides a favorable operating regime for safe and precise endovascular interventions.

As introduced earlier, during endovascular interventions, the magnetic milli-spinner first navigates downstream to the target site to perform therapeutic tasks, and then swims upstream toward the introducing sheath for safe retrieval [22]. To ensure controllable motion during both downstream navigation and upstream retrieval, the propulsion direction of the milli-spinner is always set upstream (**Figure 1a**), i.e., against the background flow, and the bidirectional motion is regulated by tuning the rotating frequency of the applied magnetic field. This frequency-based bidirectional motion control strategy is illustrated in **Figure 9d**, which presents the displacement-time curves of the optimized magnetic milli-spinner during upstream and downstream motion under a 60 beats/min pulsatile flow with a peak velocity of 40 cm/s and an average velocity of 20 cm/s. When the rotating magnetic field is set to 130 Hz, the spinning-induced propulsion velocity exceeds the average flow velocity, enabling the milli-spinner to swim upstream against the flow at $v_{avg}$ = 2.4 cm/s (**Figure 9d-i**). In contrast, when the rotating frequency is reduced to 100 Hz, the propulsion velocity generated by rotation becomes lower than the average flow velocity, and the milli-spinner is carried downstream by the flow at $v_{avg}$ = −3.0 cm/s (**Figure 9d-ii** and **Movie S7**). In this manner, bidirectional motion with controllable velocity is achieved solely by tuning the rotating frequency of the magnetic field.

It should be stated that, in clinically relevant vasculatures, vessels often exhibit complex geometries, including branching and curvature. In such scenarios, controllable and precise navigation can be achieved by integrating the magnetic actuation system with a six-axis robotic arm carrying a rotating permanent magnet. This setup enables full 3D control of the magnetic field direction and orientation, thereby allowing precise steering of the milli-spinner. By programming the robotic arm to follow a predefined trajectory and dynamically adjusting the orientation of the rotating magnetic field in real time, the milli-spinner can be guided through curved and branching vascular pathways. This multi-degree-of-freedom magnetic control strategy has been demonstrated in our previous work [22] and has also been widely adopted in magnetic microrobots in the literature [16, 21], providing a feasible and effective approach for navigation in complex physiological vessel networks.

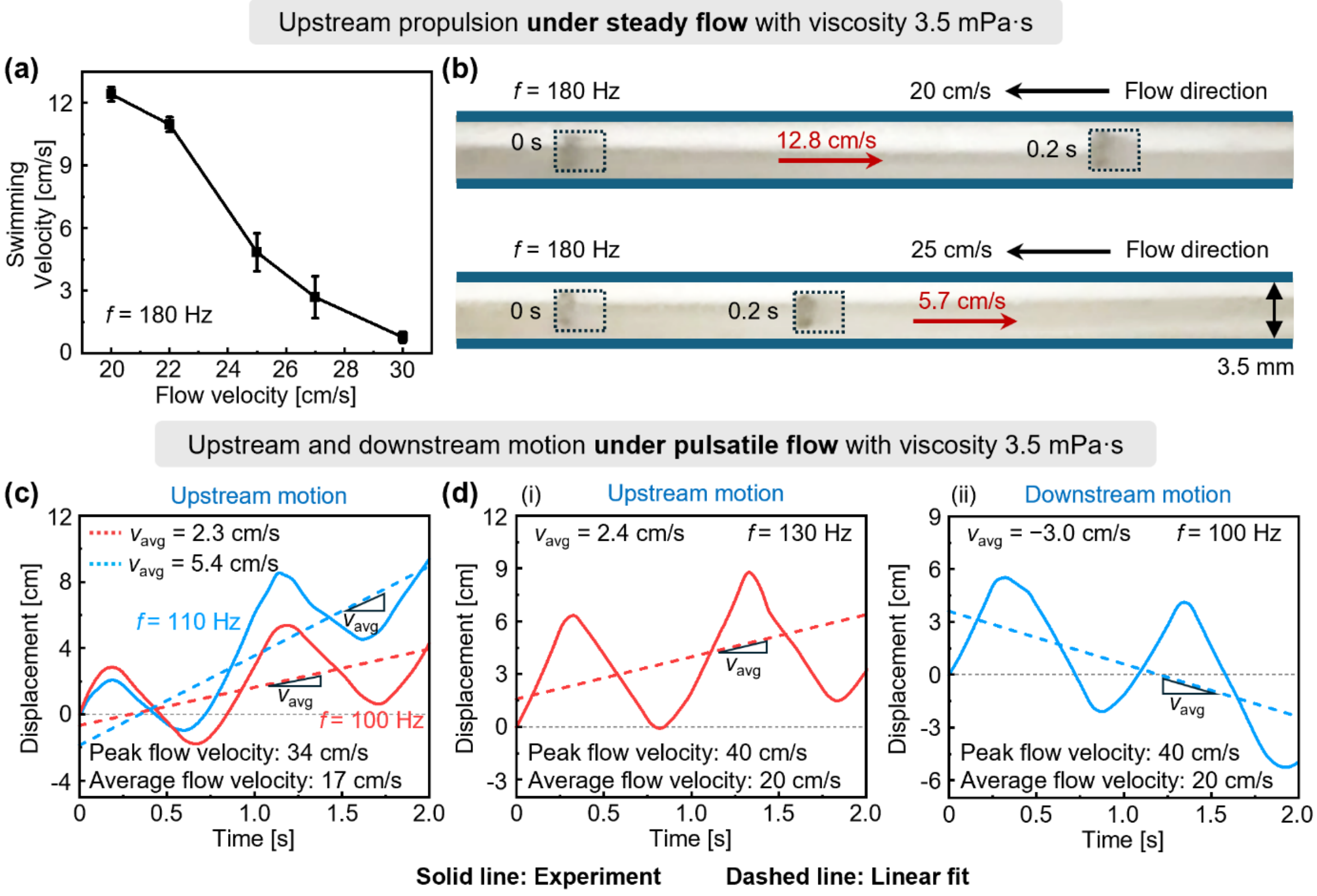

**Figure 9.** Swimming performance of the optimized magnetic milli-spinner in a 3.5 mm-ID tube filled with a saline water-glycerin mixture (viscosity 3.5 mPa·s) under steady and pulsatile flows. (a) Absolute upstream swimming velocity of the milli-spinner at $f$ = 180 Hz under steady flows with different flow velocities. (b) Experimental demonstrations of the upstream propulsion of the milli-spinner under steady background flows with flow velocities of 20 cm/s and 25 cm/s. (c) Displacement-time curves of the upstream propulsion of the milli-spinner for achieving different prescribed average upstream swimming velocities under pulsatile flow with a peak velocity of 34 cm/s and an average velocity of 17 cm/s. (d) Displacement-time curves corresponding to the (i) upstream and (ii) downstream motion of the milli-spinner under pulsatile flow with a peak velocity of 40 cm/s and an average velocity of 20 cm/s.

## 3. Conclusion

In summary, we performed design optimization for magnetic milli-spinners featuring a cylindrical body with a central through-hole, side slits, and helical fins through a combination of computational fluid dynamics simulations and experimental validation. To identify the optimized design, we systematically investigated the effects of several key parameters, including through-hole radius, fin number, fin helical angle, and slit dimension, on the propulsion performance of magnetic milli-spinners in a tubular environment under a rotating magnetic field. Both simulations and experiments demonstrate that these parameters play critical roles in enhancing the propulsion velocity and internal pressure drop of magnetic milli-spinners. By rationally selecting these parameters, an optimized milli-spinner design is proposed, which achieves propulsion velocities of 55 cm/s (~175 body lengths/s) in saline water and 44 cm/s (~140 body lengths/s) in a fluid with a viscosity of 3.5 mPa·s (comparable to that of arterial blood at high shear rates [25]) under a rotating magnetic field of 180 Hz and 20 mT. These velocities substantially exceed those of existing untethered magnetic robots in tubular environments (< 80 body lengths/s) [18-22]. Additional experiments are conducted to further evaluate the swimming performance of the optimized milli-spinner under clinically relevant flow conditions, and results indicate that the milli-spinner can stably swim upstream against strong physiological flows representative of major arteries and veins. Such high propulsion efficiency enables wireless operation in tortuous vascular environments with high flow rates, opening new opportunities for magnetic milli-spinners in a broad range of endovascular procedures, including robotic mechanical thrombectomy, embolectomy, and targeted drug delivery.

## 4. Materials and methods

### 4.1. Computational fluid dynamics simulations

To qualitatively evaluate the propulsion performance of magnetic milli-spinners with various design parameters in tubular environments, CFD simulations are conducted using COMSOL Multiphysics 6.1 (COMSOL Inc., USA). The Reynolds number in the tubular system is defined as $Re = \rho UD/\mu$ [33], where $U = \pi fD$ is the characteristic velocity, with $f$ and $D$ representing the rotating frequency and the OD of the milli-spinner, respectively, $\rho$ is the fluid density, and $\mu$ is the dynamic viscosity. The densities and viscosities of the fluids considered in this study are provided in **Table S5** in the Supporting Information. The highest rotating frequency considered in the simulation is 200 Hz, corresponding to Reynolds numbers of 3927 for saline water with a viscosity of 1.0 mPa·s and 1210 for a saline water-glycerin mixture with a viscosity of 3.5 mPa·s, which are both below the turbulent regime (typically $Re$ > 4000). As a result, the laminar flow model combined with a frozen motor method is employed in all simulations. In the frozen rotor method, a cylindrical rotating domain, slightly larger than the magnetic milli-spinner, is defined, with a rotational velocity prescribed along its longitudinal axis. Moreover, zero-pressure boundary conditions are applied at both the inlet and outlet, and no-slip wall conditions are imposed on the milli-spinner surface and tube wall. To minimize the influence of boundary effects, a 300 mm long tube filled with saline water or a saline water-glycerin mixture is used as the computational flow domain. Stationary analysis is performed to determine the steady-state flow field, pressure profile, and the resulting forces acting on the milli-spinner. In particular, the axial force on the magnetic milli-spinner is calculated by integrating the stress in the longitudinal direction over its surface. To estimate the equilibrium propulsion velocity generated by this force, an initial velocity is specified for the tube wall and iteratively adjusted until the milli-spinner reaches equilibrium. In the equilibrium state, the computed axial force approaches zero (values below $10^{-6}$ N are considered negligible), indicating that the prescribed tube wall velocity corresponds to the magnitude of the milli-spinner's equilibrium propulsion velocity but in the opposite direction.

### 4.2. Experiments

The CFD-predicted propulsion velocities of magnetic milli-spinners are validated experimentally. In the experiments, the milli-spinner body is fabricated using a customized digital light processing (DLP) 3D printer, with Anycubic ABS-Like Resin 3.0 as the printing material. A ring-shaped neodymium magnet (grade N50, SM Magnetics, USA) with an ID of 1 mm, OD of 2 mm, and length of 1 mm is attached to one end of the milli-spinner using adhesive. When subjected to a rotating magnetic field, the ring magnet drives the milli-spinner to rotate, thereby generating propulsion in the fluid. The rotating magnetic field is produced by a three-axis Helmholtz coil system (see **Section S2** and **Figure S9** in the Supporting Information for details of the experimental setup and magnetic actuation). Propulsion tests are conducted in a tube (3.5 mm ID, 6.0 mm OD, and 100 mm long) filled with saline water (**Figures 2-7 and 8c,d**) or a saline water-glycerin mixture (**Figures 8b,c and 9**), positioned within the working space of the Helmholtz coil system. For the tests with background flow (**Figure 9**), the tube's two ends are connected to a pump to generate steady or pulsatile flow. The magnetic milli-spinner is initially placed at one end of the tube and swims to the other endupon application of a rotating magnetic field. To achieve stable propulsion, a magnetic field strength of 15 mT is employed for rotating frequencies $f \leq 160$ Hz, whereas 20 mT is used at frequencies $f > 160$ Hz. The entire propulsion process is recorded using a camera, from which the displacement-time curve is extracted. For tests conducted without background flow or under steady flow, the slope of the linear portion of this curve is taken as the equilibrium swimming velocity of the magnetic milli-spinner. For tests under pulsatile flow, a linear fit to the displacement-time curve is used to determine the average swimming velocity.

**Acknowledgment**

Lu would like to thank Dr. Shuai Wu at Oregon State University for helpful discussions.

**Conflicts of Interest**

The authors declare no conflicts of interest.